\documentclass[aps,twocolumn,amsmath,amssymb,floatfix,showpacs,
superscriptaddress]{revtex4}

\usepackage[dvips]{graphics}
\usepackage{color}
\definecolor{blue}{rgb}{0,0,0.7}
\definecolor{red}{rgb}{1,0,0}

\begin{document}

\title{\textcolor{blue}{Effect of Dephasing on Electron Transport in 
a Molecular Wire: Green's Function Approach}}

\author{Moumita Dey}

\affiliation{Theoretical Condensed Matter Physics Division, Saha 
Institute of Nuclear Physics, Sector-I, Block-AF, Bidhannagar,
Kolkata-700 064, India}

\author{Santanu K. Maiti}

\email{santanu.maiti@saha.ac.in}

\affiliation{Theoretical Condensed Matter Physics Division, Saha
Institute of Nuclear Physics, Sector-I, Block-AF, Bidhannagar,
Kolkata-700 064, India}

\affiliation{Department of Physics, Narasinha Dutt College, 129
Belilious Road, Howrah-711 101, India}

\author{S. N. karmakar}

\affiliation{Theoretical Condensed Matter Physics Division, Saha
Institute of Nuclear Physics, Sector-I, Block-AF, Bidhannagar,
Kolkata-700 064, India}

\begin{abstract}
The effect of dephasing on electron transport through a benzene molecule
is carefully examined using a phenomenological model introduced by
B\"{u}ttiker. Within a tight-binding framework all the calculations are
performed based on the Green's function formalism. We investigate the 
influence of dephasing on transmission probability and current-voltage 
characteristics for three different configurations ({\em ortho}, 
{\em meta} and {\em para}) of the molecular system depending on the 
locations of two contacting leads. The presence of dephasing provides 
a significant change in the spectral properties of the molecule and 
exhibits several interesting patterns that have so far remain unexplored.
\end{abstract}


\maketitle

\section{Introduction}

Ongoing trend of miniaturizing electronic devices eventually approaches 
the ultimate limit where even a single molecule can be used as an 
electrical circuit element. Idea of devicing a single molecule as the 
building block of future generation electronics seems fascinating 
because of the possibility to assemble a large number of molecules 
onto a chip i.e., remarkable enhancement in integration density can 
take place~\cite{ratner1}. Discovery of sophisticated molecular scale 
measurement methodologies such as scanning tunneling microscopy (STM), 
atomic force microscopy (AFM), scanning electro-chemical microscopy (SECM), 
etc. have made it possible to study electron transport phenomena in 
molecular bridge systems~\cite{chen}. Unlike conventional semiconductor 
electronic transport properties can not be investigated with Boltzmann 
transport equation as quantum coherence has a crucial significance on 
electron transport at this length scale and a full quantum mechanical 
treatment is needed~\cite{datta1,datta2}. At low temperature and small 
bias electron transport is quite successfully explained within the 
framework of Landauer formalism~\cite{land}. But, validity of Landauer 
formulation is limited within coherent transport regime where it is
assumed that electrons suffer only elastic scattering within the 
conductor. This assumption allows us to neglect any kind of inelastic 
processes, which is rather unrealistic unless the time scale related 
to transport phenomena is much faster than the nuclear motion~\cite{amato}.

A main source of dephasing is electron-phonon (e-ph) interaction in 
molecular transport junctions. Experimentally the strength
of e-ph interaction in molecular systems can be quantified through the
measurement of molecular vibrational spectrum using inelastic tunneling
spectroscopy~\cite{ho,wang,gal}. Roughly, it corresponds to finding 
locations of the peaks in the second order derivative of a current-voltage
characteristic, where the voltages at the peaks match with eigenenergies
of the phonons. Rigorous analysis of e-ph interaction 
can be done self-consistently with the help of density functional theory 
(DFT) and non-equilibrium Green's function formalism (NEGF)~\cite{guo2,guo3}. 
But this method is notably time consuming and becomes very difficult to 
do even for a molecular conductor comprising a moderate number of atoms. 
Another way out was proposed by B\"{u}ttiker over two decades 
ago~\cite{butti1,butti2}. He came up with 
an elegant idea of incorporating the effect of phase breaking due to 
inelastic scattering phenomenologically by introducing some fictitious 
voltage probes into the coherent system. While moving from source to 
drain electrons get scattered into the fictitious floating voltage probes 
and then re-emitted into the device by loosing the phase memory. To 
implement this idea each inelastic scatterer is modeled in terms of 
electron reservoir (or infinite impedance voltage probe) and coupled to 
the device. Since the net current flowing through the fictitious probes 
has to be zero to satisfy the conservation of total number of particles, 
electro-chemical potentials $(\mu)$ of each lead has to be adjusted 
accordingly. Due to its appealing simplicity B\"{u}ttiker probe model has 
been extensively used in studying quantum transport in low-dimensional 
systems.

The idea of using molecules as active components of a device was suggested 
by Aviram and Ratner over three decades ago~\cite{aviram}. Since then 
several ab-initio and model calculations have been performed to investigate 
molecular transport theoretically~\cite{ventra1,ventra2,sumit,tagami,orella1,
orella2,arai,san1,san2}. But experimental realizations took a little longer 
time to get feasible. In 1997, Reed and co-workers studied the 
current-voltage ($I$-$V$) 
characteristics of a single benzene molecule attached to electrodes via 
thiol groups~\cite{reed}. Various other experiments have been reported in 
literature exploring many interesting features e.g., ballistic transport, 
quantized conductance, negative differential resistance (NDR), gate 
controlled transistor operation to name a few. Effect of dephasing in 
molecular systems has been studied in few literatures both 
theoretically~\cite{yan1,yan2} as well as experimentally. In 1989, 
Amato and Pastawski investigated the effect of inelastic scattering 
processes on a tight-binding linear disordered chain using B\"{u}ttiker 
probe model~\cite{amato}. In 2007, Datta {\em et al.} proposed a different 
phenomenological model which provides the flexibility of adjusting degree 
of phase and momentum relaxation independently~\cite{dattadephase}. In 
2008, Nozaki {\em et al.} obtained transport properties through various 
molecular junctions by B\"{u}ttiker model using Extended H\"{u}ckel 
theory~\cite{nozaki}. In 2009, Guo and co-workers have reported an 
ab-initio calculation~\cite{guo1} including the effect of phase breaking 
in 1,4-benzenedithiol molecule attached to two Al electrodes, gold QPC and 
a very thin Al(001) nanowire. They have also done a comparative analysis 
between B\"{u}ttiker probe model and Datta dephasing 
model~\cite{dattadephase}. Though various efforts have already been made 
to explain the basic features of electron dephasing on molecular transport,
but to the best of our knowledge, no rigorous effort has been made so far 
to unravel the combined effect of quantum interference and electron
dephasing on molecular transport. This is the main motivation behind
this work.

In our present article we study two terminal electron transport through
a single benzene molecule including the effect of dephasing by a discrete 
tight-binding model. We obtain the transmission probability using Green's 
function technique within the framework of Landauer-Buttiker formulation.
Transmittance-energy and current-voltage characteristics are obtained
for three different configurations ({\em para}, {\em ortho} and {\em meta}) 
of benzene molecule depending on three different positions of the drain 
with respect to the source. The presence of dephasing provides a 
significant change in the spectral properties of the molecule and here
we essentially focus our results in this aspect. Our model calculation
can be extended further for any complicated molecular structure.

In what follows, we present the results. In section II, the model
and the theoretical formulation are presented. Section III describes 
the results, and finally, we make our conclusions in section IV.

\section{Molecular Model and Theoretical Formulation}

\begin{figure}[ht]
{\centering \resizebox*{8cm}{4.5cm}{\includegraphics{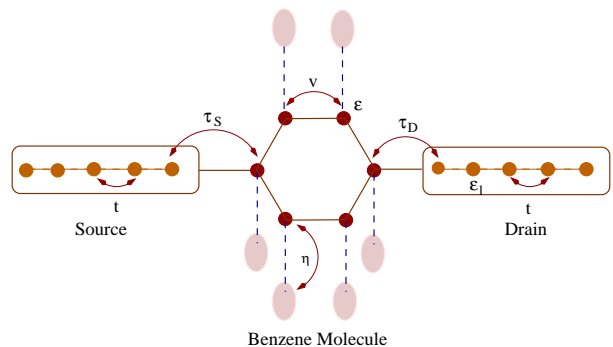}}\par}
\caption{(Color online). Schematic diagram of a single benzene molecule 
attached to two leads, namely, source and drain. Each molecular site is
assumed to be coupled to a fictitious electron reservoir, shown by the
pink shaded area, which takes into account the effect of phase breaking 
processes.}
\label{dephasing}
\end{figure}

We start by describing our model, illustrated in Fig.~\ref{dephasing}, 
where a single benzene molecule is connected symmetrically to two 
one-dimensional, semi-infinite leads, commonly known as source and drain.
The leads are characterized by the electrochemical potentials $\mu_S$ 
and $\mu_D$, respectively, under the non-equilibrium condition when an 
external bias voltage is applied. Both the two leads and the molecule 
are simulated by a simple tight-binding Hamiltonian within 
nearest-neighbor hopping approximation. Following B\"{u}ttiker's idea, 
the effect of dephasing due to inelastic scattering is incorporated in 
the model phenomenologically by coupling each spatial sites to a different 
fictitious electron reservoir or in other words an infinite impedance 
voltage probe (shown by the pink shaded area in Fig.~\ref{dephasing}) 
via perfect leads. The key idea is that {\em the fictitious floating probes 
extract electrons from the device to the reservoir, but as the net current 
flowing through these probes is zero, so they re-inject the electrons into 
the conductor after phase randomization and thus effectively play the role 
of phase breaking scatterers.}

The Hamiltonian for the entire system is given by the sum of four terms
\begin{equation}
H= H_{mol} + H_{leads} + H_{tun} + H_{dephase}.
\label{eqn1}
\end{equation}
The first term represents the Hamiltonian for the single benzene molecule
consisting of six sites ($N=6$) which is coupled to two electron reservoirs 
through conducting leads, namely, source and drain. The molecule is modeled 
by the nearest-neighbor tight-binding Hamiltonian which in Wannier basis 
can be written as, 
\begin{equation}
H_{mol} = \sum_i \epsilon d_{i}^{\dag}d_i + \sum_i v [d_{i+1}^{\dag}d_{i} 
+ h.c.],
\label{eqn2}
\end{equation}
where $\epsilon$ refers to the site energy of an electron at the 
$i$-th site of the molecular system and $v$ represents the isotropic 
nearest-neighbor coupling strength between the molecular sites. 
$d_{i}^{\dag}$ and $d_i$ correspond to the creation and annihilation 
operators, respectively, of an electron at the $i$-th site of the 
molecule.

Similarly the second and third terms of Eq.~(\ref{eqn1}) denote the 
Hamiltonians for the one-dimensional semi-infinite leads (source and 
drain) and molecule-to-lead coupling. In Wannier basis representation 
they can be written as follows.
\begin{eqnarray}
H_{leads} & = & H_S + H_D \nonumber \\
& = & \sum_{\alpha=S,D} \left\{\sum \limits_{n}\epsilon_{l}
c_{n}^{\dag}c_n  + \sum_{n} t[c^{\dag}_{n+1} c_n + h.c.]\right\},
\nonumber \\
\label{eqn3}
\end{eqnarray}
and,
\begin{eqnarray}
H_{tun} & = & H_{S,mol} + H_{D,mol} \nonumber \\
& = &   \tau_S[d_1^{\dag}c_0 + h.c.] + \tau_D[d_p^{\dag}c_{N+1} + h.c.].
\label{eqn4}
\end{eqnarray}
Here, $\epsilon_l$ and $v$ stand for the site energy and nearest-neighbor
hopping between the sites of the two leads. $c_{n}^{\dag}$ and $c_{n}$ are 
the creation and annihilation operators, respectively, of an electron at the 
site $n$ of the leads. The hopping integral between the source and the 
molecule is $\tau_S$, while it is $\tau_D$ between the molecule and the 
drain. Usually benzene molecules are attached to gold leads via thiol 
groups (-SH bond) in the chemisorption technique. Here, the effect of 
such substituent (e.g., thiol group) is included in the parameters 
$\tau_S$ and $\tau_D$. The source is assumed to be connected at the site 
number $1$, while the drain is coupled to the $p$-th site, which is 
variable, of the molecular system. The integer $p$ becomes $2$, $3$ and
$4$ for the {\em ortho}, {\em meta} and {\em para} configurations,
respectively.

In order to incorporate the effect of dephasing, external electron 
reservoirs are phenomenologically introduced and they are connected to all 
the molecular sites (six) through fictitious probes. The phase breaking 
probes are identically represented within tight-binding framework like 
source and drain as given below.
\begin{equation}
H_{dephase}=\sum_{i=1}^{N}\left[H_{i}^{d,probes}+H_{i}^{d,tun}\right],
\label{eqn5}
\end{equation}
where,
\begin{equation}
H_{i}^{d,probes}=\sum_{m} \epsilon_{d} c_{m}^{D i \dag}c_m^{D i}  
+ \sum_m t_d[c^{D i \dag}_{m+1} c_m^{D i} + h.c.],
\label{eqn6}
\end{equation}
and,
\begin{equation}
H_{i}^{d,tun} = \eta \left[ c^{D i \dag}_{0}d_{i} + h.c.\right].
\label{eqn7}
\end{equation}
Here, $\epsilon_d$ and $t_d$ being the site energy and the hopping strength
between the nearest-neighbor sites of the dephasing probes, respectively. 
$c^{D i \dag}_{m} (c^{D i}_{m})$ is the creation (annihilation) operator of
an electron at the $m$-th site of the $i$-th lead. The dephasing strength 
is characterized by the coupling parameter $\eta$.

To obtain the transmission probability of an electron through such a
molecular bridge system, we use Green's function formalism. Within the 
regime of coherent transport and in the absence of Coulomb interaction 
this technique is well applied.

The single particle Green's function operator representing the entire 
system for an electron with energy $E$ is defined as,
\begin{equation}
G=\left( E - H + i\eta \right)^{-1}
\label{eqn8}
\end{equation}
where, $\eta \rightarrow 0^+$.

Following the matrix form of \mbox{\boldmath $H$} and \mbox{\boldmath $G$} 
the problem of finding \mbox{\boldmath $G$} in the full Hilbert space 
\mbox{\boldmath $H$} can be mapped exactly to a Green's function 
\mbox{\boldmath $G_{mol}^{eff}$} corresponding to an effective 
Hamiltonian in the reduced Hilbert space of the molecule itself and 
we have,
\begin{equation}
\mbox{\boldmath ${\mathcal G}$=$G_{mol}^{eff}$}=\left(\mbox {\boldmath $E- 
H_{mol}-\Sigma_S-\Sigma_D-\sum\limits_{k=1}^{N} \Sigma_k $}\right)^{-1},
\label{equ9}
\end{equation}
where, 
\begin{eqnarray}
\mbox{\boldmath $\Sigma_{S(D)}$} & = & \mbox{\boldmath 
$H_{S,mol(mol,R)}^{\dag} G_{S(D)} H_{S,mol(mol,R)}$}, \nonumber \\
\mbox{\boldmath $\Sigma_{k}$} & = & \mbox{\boldmath 
$H_{k}^{d,tun \dag} G_{k} H_{k}^{d,tun}$}. 
\label{eqn10}
\end{eqnarray}
These \mbox{\boldmath $\Sigma_{S(D)}$} and \mbox{\boldmath $\Sigma_{k}$} are
the contact self-energies introduced to incorporate the effect of coupling
of the molecule to the source (drain) and the fictitious electron reservoirs.
It is evident from Eq.~(\ref{eqn10}) that the form of the self-energies are 
independent of the molecule itself through which transmission is studied.
Using Dyson equation the analytic form of the self energies can be evaluated
as follows,
\begin{equation}
\Sigma_{\nu} = \frac{\tau_{\nu}^2}{E - \epsilon_{\nu}-\sigma_{\nu}}.
\label{eqn11}
\end{equation}
Here, $\sigma_{\nu} = (E-\epsilon_{\nu})/2 - i\sqrt{t_{\nu}^2-
(E-\epsilon_{\nu})^2/4}$; and $\nu = S,D,1,2, \dots N$. For source and 
drain, $\nu=S(D)$, $\tau_{\nu}=\tau_{S(D)}$, $\epsilon_{\nu}=\epsilon $, 
and $t_{\nu}=t$; while considering the dephasing probes, 
$\nu=k~(k=1,2, \dots N)$, $\tau_{\nu}=\eta$, $\epsilon_{\nu}=\epsilon_l$, 
and $t_{\nu}=t_d$.

Following D'Amato-Pastawski model~\cite{amato} the effective transmission 
probability of an electron from source to drain including the partially 
phase breaking process is given by the following expression.
\begin{equation}
T_{eff} = T_{S,D} + \sum \limits_{i,j=1}^{N} \mbox{\boldmath$T$}_{D,i}~
\mbox {\boldmath $W$}^{-1}_{i,j}~\mbox{\boldmath $T$}_{j,S}.
\label{eqn12}
\end{equation}
The above expression has a significant physical implication. The first 
term is the coherent contribution to electron transmission, whereas the 
second term refers to the incoherent component of tunneling due to 
electrons suffering phase randomizing scattering at the fictitious 
dephasing reservoirs. $T_{i,j}$ is the transmission probability between 
any pair of reservoirs {$i,j$}$~(i,j=S,D,1,2 \dots N)$ and it is 
expressed as follows,
\begin{equation}
T_{i,j} = \mbox{Tr}\mbox{\boldmath [$\Gamma_{i} \mathcal {G}^r \Gamma_{j} 
\mathcal {G}^a$]}.
\label{eqn13}
\end{equation}
$\Gamma_{i}$'s are the coupling matrices representing the coupling between 
the molecule and the leads and they are mathematically defined by the 
relation,
\begin{equation}
\mbox {\boldmath $\Gamma_i$} = i \left[\mbox {\boldmath $\Sigma^r_{i} - 
\Sigma^a_{i}$}\right]
\label{eqn14}
\end{equation}
Here, \mbox{\boldmath $\Sigma_{k}^r$} and \mbox{\boldmath $\Sigma_{k}^a$}
are the retarded and advanced self-energies associated with the $k$-th
lead, respectively.

In the second term of Eq.~(\ref{eqn12}), \mbox{\boldmath $W$} matrix is
defined by the following relation,
\begin{equation}
W_{i,j} = [(1-R_{i,i})\delta_{ij} - T_{i,j}(1 - \delta_{ij})],
\label{eqn15}
\end{equation}
where,
\begin{equation}
R_{i,i} = 1 - \sum \limits_{j \neq i} T_{i,j}.
\label{eqn16}
\end{equation}
$R_{i,i}$ is the reflection probability of an electron from $i$-th lead.

It is shown in literature by Datta {\em et al.}~\cite{datta1,datta2} that the 
self-energy can be expressed as a linear combination of a real and an
imaginary part in the form,
\begin{equation}
\mbox{\boldmath ${\Sigma^r_{j}}$} = \mbox{\boldmath $\Lambda_{j}$} - 
i \mbox{\boldmath $\Delta_{j}$}.
\label{equ17}
\end{equation}
The real part of self-energy describes the shift of the energy levels
and the imaginary part corresponds to the broadening of the levels. The
finite imaginary part appears due to incorporation of the semi-infinite
leads having continuous energy spectrum. Therefore, the coupling matrices
can easily be obtained from the self-energy expression and is expressed as,
\begin{equation}
\mbox{\boldmath $\Gamma_{i}$}=-2\,{\mbox {Im}} 
(\mbox{\boldmath $\Sigma_{i}$}).
\label{equ18}
\end{equation}
Considering linear transport regime, at absolute zero temperature the 
linear conductance $(g_{eff})$ is obtained using two-terminal Landauer 
conductance formula,
\begin{equation}
g_{eff} =\frac{2 e^2}{h}T_{eff}(E_F).
\label{equ19}
\end{equation}
With the knowledge of the effective transmission probability we compute 
the current-voltage ($I$-$V$) characteristics by the standard formalism 
based on quantum scattering theory.
\begin{equation}
I(V) = \frac{2 e}{h} \int \limits_{- \infty}^{\infty} 
T_{eff}[f_S(E)-f_D(E)]\,dE.
\label{eqn20}
\end{equation}
Here, $f_{S(D)}(E) = \left[1+e^{\frac{E-\mu_{S(D)}}{k_{B}T}} \right]^{-1}$ 
is the Fermi function corresponding to the source and drain. At absolute
zero temperature the above equation boils down to the following expression.
\begin{equation}
I(V) = \frac{2 e}{h} \int \limits_{E_F - \frac{eV}{2}}^{E_F + 
\frac{eV}{2}} T_{eff} \,dE.
\label{eqn21}
\end{equation}
In our present work we use the above expression assuming that the 
potential drop takes place only at the boundary of the molecular system. 
This assumption is good enough for molecules of smaller size. Throughout 
our study we choose $c=e=h=1$ for the sake of simplicity.

\section{Numerical results and discussion}

In this section we present the results obtained by numerical simulation 
considering a single benzene molecule sandwiched between source and 
drain. The orbital energies $(\epsilon)$ of the molecule are set to 
zero and the nearest-neighbor hopping matrix element $(t)$ is fixed 
at $2.4$\,eV~{\textcolor{red}{\cite{sumit}}} for 
an aromatic ring structure of the benzene molecule. The other hopping 
parameters in the leads $(t)$ and the dephasing probes $(t_{d})$ are 
fixed at $4$\,eV, while the site energies are chosen to be zero i.e., 
$\epsilon_l = \epsilon_d = 0$. The dephasing strength, characterized 
by the parameter $\eta$, is fixed
\begin{figure}[ht]
{\centering \resizebox*{7.5cm}{4cm}{\includegraphics{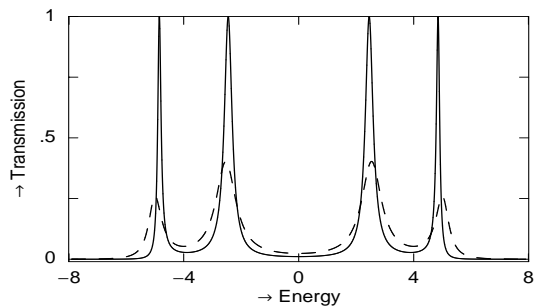}}\par}
\caption{Variation of transmission probability as a function of energy for 
a benzene molecule where the drain is connected at the {\em para} position 
with respect to the source. The solid and dashed curves 
correspond to the results in the absence and presence of dephasing, 
respectively.}
\label{transpara}
\end{figure}
at $1$\,eV. Throughout our analysis the coupling strengths of the molecule 
to the source and drain, characterized by the parameters $\tau_S$ and 
$\tau_D$, are set at $1$ eV. The molecular coupling to the contacting 
leads is one of the most important factors which regulates the electronic 
transmission through a molecular wire and this phenomenon has been 
\begin{figure}[ht]
{\centering \resizebox*{7.5cm}{4cm}
{\includegraphics{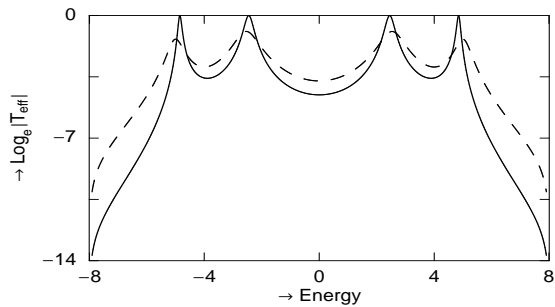}}\par}
\caption{The results of Fig.~\ref{transpara} are re-plotted in a 
logarithmic scale to clarify the effect of dephasing more clearly.}
\label{transparanew}
\end{figure}
extensively studied in our previous works~\cite{san1,san2}. Accordingly,
here we describe all the essential features of electron transport for a 
particular molecular coupling strength. The molecular transport properties 
are also highly sensitive on the molecule-to-lead interface geometry, and, 
in the present work we illustrate how they are influenced in the presence 
of electron dephasing. To reveal these facts first we discuss the behavior 
of transmittance-energy characteristics and then we describe the nature 
of current-voltage spectra for three different configurations ({\em ortho}, 
{\em meta} and {\em para}) of the benzene molecule.

Here, we establish all the results at absolute zero temperature and 
restrict ourselves within elastic dephasing so that no energy is exchanged 
between the transported electrons and the external fictitious reservoirs, 
only the phase informations get lost. This condition allows the incoherent 
electrons to generate a steady-state current through the sample.

\subsection{Transmittance-energy characteristics}

\subsubsection{Molecule coupled symmetrically}

In Fig.~\ref{transpara} we show the variation of transmission probability 
as a function of injecting electron energy for a benzene molecule, where 
the molecule is coupled symmetrically to the source and drain i.e., the 
\begin{figure}[ht]
{\centering \resizebox*{7.5cm}{4cm}{\includegraphics{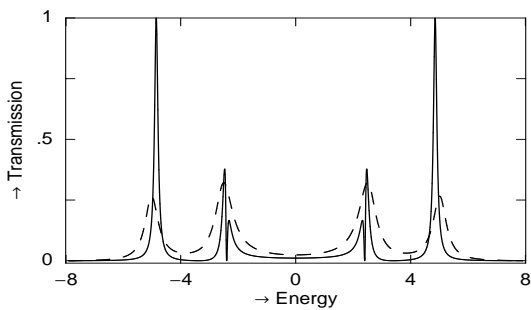}}\par}
\caption{Transmission probability versus energy characteristics of a 
benzene molecule where the drain is connected at the {\em ortho} position 
with respect to the source. The solid and dashed lines represent the 
results in the absence and presence of dephasing, respectively.}
\label{transortho}
\end{figure}
upper and lower arms of the molecular ring have identical length. This is 
the so-called {\em para} configuration. The solid curve
\begin{figure}[ht]
{\centering \resizebox*{7.5cm}{4cm}{\includegraphics{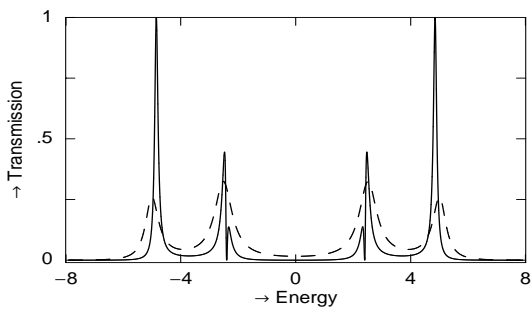}}\par}
\caption{Transmission probability versus energy characteristics of a benzene 
molecule where the drain is coupled to the {\em meta} position with respect 
to the source. The solid and dashed curves correspond to the results in 
the absence and presence of dephasing, respectively.}
\label{transmeta}
\end{figure}
represents the variation of coherent transmission probability, while the 
dashed curve depicts the result in the presence of
electron dephasing. It is observed that, in the absence of dephasing the
transmission probability exhibits sharp resonant peaks (see solid line) 
for some specific energies, whereas it almost vanishes for 
all other energy values. At these resonances, the transmission probability 
$T_{eff}$ 
goes to unity. All the resonant peaks are associated with the energy 
eigenvalues of the benzene molecule, and therefore, we can say the 
transmittance spectrum is a fingerprint of the electronic structure of 
the molecule. The situation becomes much interesting when the effect of 
dephasing is introduced. It shows that the magnitude of the resonant 
\begin{figure}[ht]
{\centering \resizebox*{7.5cm}{4cm}
{\includegraphics{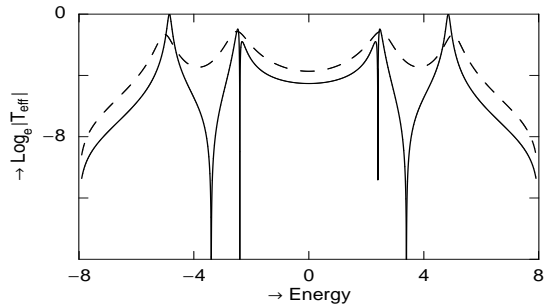}}\par}
\caption{The results of Fig.~\ref{transortho} are re-plotted in 
a `log' scale.}
\label{transorthonew}
\end{figure}
peaks gets suppressed enormously (dashed line) compared to the coherent 
transmission (solid line). This is due to the increased rate of scattering 
in the additional fictitious 
\begin{figure}[ht]
{\centering \resizebox*{7.5cm}{4cm}
{\includegraphics{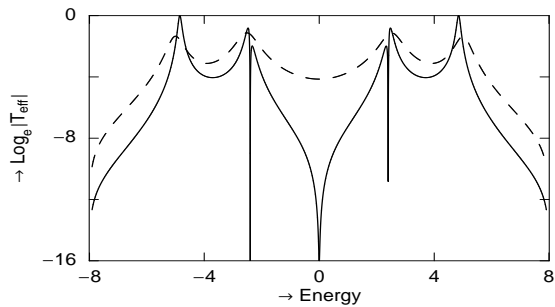}}\par}
\caption{The results of Fig.~\ref{transmeta} are re-plotted in 
a `log' scale.}
\label{transmetanew}
\end{figure}
electron reservoirs. Additionally, we also notice that the widths of the 
resonant peaks become broadened. This broadening is caused by the 
dominance of phase randomizing effect over backscattering due to loss
of phase coherence of interfering electrons in the presence of dephasing 
reservoirs. To judge the effect of dephasing more
transparently, in Fig.~\ref{transparanew} we re-plot the results
of Fig.~\ref{transpara} in a logarithmic scale, where the solid and dashed
lines represent the identical meaning as above. Since the transmission 
probability lies within the range $0$ to $1$, it becomes zero or negative
in the logarithmic scale as shown in Fig.~\ref{transparanew}. It is observed
that the variations in the transmission peaks between the two curves (solid
and dashed) get suppressed, but the broadening effect in the presence of 
dephasing can be much clearly observed from the `log' scale.

\subsubsection{Molecule coupled asymmetrically}

A significant change in the transmission spectrum is observed when the
benzene molecule is coupled asymmetrically to the source and drain 
i.e., the upper and lower arms of the molecular ring have unequal 
lengths. For the {\em ortho} configuration the results are given in 
Fig.~\ref{transortho}, while in Fig.~\ref{transmeta} the results are 
shown for the {\em meta} configuration. From the spectra 
(Figs.~\ref{transortho} and \ref{transmeta}) we notice that in the 
absence of dephasing some resonant peaks do not reach to unity and 
their amplitudes are significantly reduced compared to the other resonant
peaks. This is solely due to the effect of quantum interference between
the electronic waves passing through the upper and lower arms of the
molecular ring. Quite interestingly we observe that in these asymmetric
molecular wires ({\em ortho} and {\em meta}) two anti-resonant states
appear in the transmittance spectrum where transmission probability drops
exactly to zero. These anti-resonant states are specific to the 
\begin{figure}[ht]
{\centering \resizebox*{7.5cm}{4cm}{\includegraphics{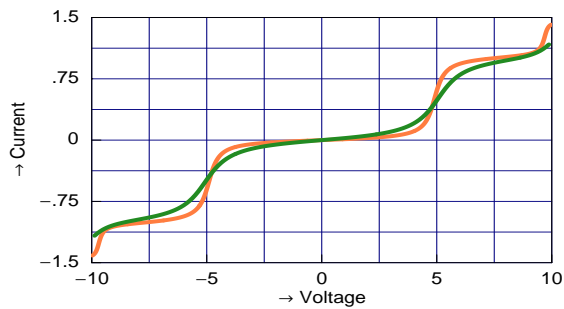}}\par}
\caption{(Color online). $I$-$V$ characteristics of a benzene molecule where 
the drain is connected at the {\em para} position with respect to the source. 
The orange and green curves correspond to the currents in the absence 
and presence of dephasing, respectively.}
\label{currpara}
\end{figure}
interferometric nature of the molecular system and their positions in the
energy scale are independent of the molecule-to-lead coupling strength.
A careful observation reveals that a sharp peak is followed by a sharp 
dip and vice versa across the anti-resonant energy which is quite analogous 
to a Fano-like line shape. All these anti-resonant states disappear as
long as dephasing of electrons is taken into account those are clearly 
observed from the dashed lines of Figs.~\ref{transortho} 
and \ref{transmeta}. Here also the magnitudes of the resonant peaks get 
decreased and they become broadened as we describe earlier in the 
{\em para} configuration. The results of 
Figs.~\ref{transortho} and \ref{transmeta} are re-drawn in 
Figs.~\ref{transorthonew} and \ref{transmetanew}, respectively, where the
transmission probabilities are determined in the `log' scale. From these
spectra (Figs.~\ref{transorthonew} and \ref{transmetanew}) the characteristic
features of anti-resonant states in the case of coherent transmission and 
the broadening of resonant peaks in the presence of electron dephasing are
clearly noticed. Our results also predict that in the asymmetric molecular
ring the positions of the anti-resonant states are strongly sensitive on
the location of measuring electrodes. Broadening of transmission peaks in
presence of dephasing is much pronounced in the case where the drain is
connected at the {\em meta} position of the benzene molecule rather than 
the {\em ortho} position, and, it can be clearly observed from the 
$Log_e T_{eff}$ vs. $E$ plot instead of the transmittance-energy spectra.

\subsection{Current-voltage characteristics}

All the above features of electron transmission become much clear from 
our study of current-voltage ($I$-$V$) characteristics. The current 
through the molecular system is obtained by integrating the transmission 
function following Eq.~(\ref{eqn21}). Here we fix the Fermi energy $E_F$ 
at zero.

\subsubsection{Molecule coupled symmetrically}

In Fig.~\ref{currpara} we plot the variation of current as a function of
applied bias voltage for a benzene molecule where the leads are connected
symmetrically. The orange and green lines represent the currents in the
absence and presence of electron dephasing. The current varies quite
\begin{figure}[ht]
{\centering \resizebox*{7.5cm}{4cm}{\includegraphics{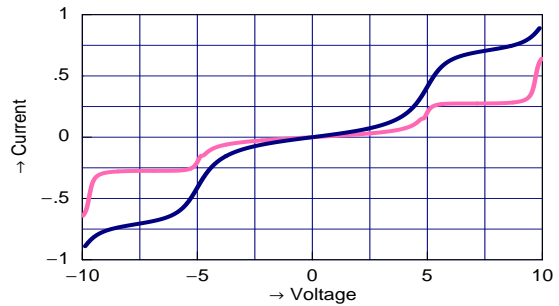}}\par}
\caption{(Color online). $I$-$V$ characteristics of a benzene molecule 
where the drain is connected at the {\em ortho} position with respect 
to the source. The pink and deep blue lines represent the currents in the 
absence and presence of dephasing, respectively. }
\label{currortho}
\end{figure}
continuously with the applied bias voltage $V$. Depending on the molecular
coupling to the side attached leads, the current shows continuous-like
or step-like behavior~\cite{san1}. Tuning the molecular coupling strength
the current amplitude through the molecular wire can be regulated nicely 
for a fixed bias voltage. This provides an interesting behavior in designing
molecular electronic devices. Both in the absence and presence of electron
dephasing current shows almost identical variation though a significant
change is observed in their conductance spectra. This is due to the fact 
that the dephasing broadens the transmission peaks, while it also suppresses 
the magnitude. These two effects nullify each other showing almost identical
spectrum in the $I$-$V$ characteristics.

\subsubsection{Molecule coupled asymmetrically}

\begin{figure}[ht]
{\centering \resizebox*{7.5cm}{4cm}{\includegraphics{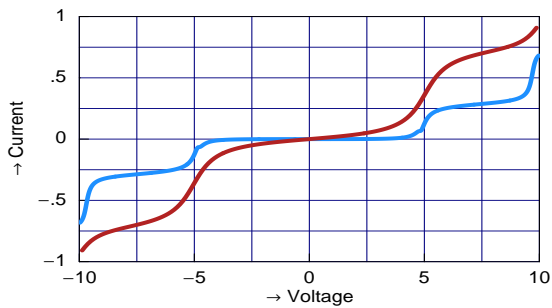}}\par}
\caption{(Color online). $I$-$V$ characteristics of a benzene molecule 
where the drain is connected at the {\em meta} position with respect to 
the source. The light blue and red lines correspond to the currents in 
the absence and presence of dephasing, respectively. }
\label{currmeta}
\end{figure}

The effect of dephasing becomes much prominent in the current-voltage 
characteristics when the molecule is coupled asymmetrically to the source
and drain. For the {\em ortho} configuration, the currents are shown in
Fig.~\ref{currortho}, while in Fig.~\ref{currmeta} the results are shown
when the molecule is coupled to the leads in the {\em meta} configuration.
Quite similar to the {\em para} configuration (Fig.~\ref{currpara}), in 
these two cases ({\em ortho} and {\em meta} configurations) also the 
current varies almost continuously with the applied bias voltage $V$.
In these asymmetric configurations, the current amplitude gets enhanced
significantly when the phase breaking effect is considered (deep blue
curve in Fig.~\ref{currortho} and light blue curve in Fig.~\ref{currmeta})
compared to the case where the molecule is free from any phase 
randomizing process (pink line in Fig.~\ref{currortho} and red line in
Fig.~\ref{currmeta}). Such an enhancement in current amplitude takes 
place because of notable broadening of the transmission peaks due to 
reduced destructive interference for dominance of phase decoherence 
over the backscattering effect. 

\section{Closing remarks}

To summarize, in the present work we have examined electron transport 
phenomena through a single benzene molecule including the effect of 
phase breaking using a phenomenological model introduced by B\"{u}ttiker.
Experimentally the benzene molecule is usually coupled to noble metal
electrodes (e.g., gold leads) through thiol linking groups. In our 
theoretical work, we present the entire system (source-molecule-drain)
by a discrete lattice model within a simple one-electron, tight-binding
framework. We have studied transmission-energy spectrum and current-voltage
characteristics considering three different configurations of the system
depending on the position of the drain contact ({\em ortho}, {\em meta} 
and {\em para} positions). We have observed that the sharp transmission 
peaks are broadened and their magnitude is also suppressed in the presence 
of dephasing due to loss of phase coherence and thus it provides an 
enhancement in the current amplitude. But the broadening effect and so as 
the increase in current amplitude is much pronounced in the case where 
the drain is connected asymmetrically to the molecule with respect to 
the source. Another significant feature is that the two anti-resonant 
states, observed in the transmittance spectrum of an asymmetrically 
connected benzene molecule, completely disappear in the presence of 
strong dephasing due to reduced destructive interference. Thus the 
effect of dephasing can provide a significant change in the 
current-voltage relation.

In the B\"{u}ttiker probe model fictitious voltage 
probes are introduced into the coherent system. These virtual probes 
introduce additional resistance in the system by relaxing the momentum 
of the electrons by partially reflecting them i.e., back-scattering. 
This model is
appropriate for those molecular systems where the dominant scattering
source does not conserve electron momentum, for example, electron-phonon
scattering which is very important in molecular conductor in finite
temperature. However, B\"{u}ttiker model does not consider the effect
of momentum relaxation. In experimental situations dephasing may also
arise from electron-electron scattering which destroy phase but not 
momentum~\cite{dattadephase,guo1}. Datta model presents the flexibility 
of tuning phase and momentum independently of each other and we intend 
to study this phenomenological model in near future.

Recent experimental progresses have significantly 
inspired in model calculations and numerical simulations to study
vibronic effects on electron transport in molecular junctions~\cite{vib1,
vib2,vib3,vib4}. Most of the studies of vibronic effects in the resonant
tunneling regime (for higher voltages) have illustrated that vibrational
motion may affect the current-voltage spectra quite significantly. But,
the point is that in the resonant tunneling region vibronic effect becomes
significant only when the molecule is coupled to the measuring leads
{\em weakly}. In a very recent work~\cite{dom1} Benesch {\em et al.}
have studied vibronic effects on conductance in molecular junctions where 
a benzene ring is coupled {\em weakly} to two gold electrodes and shown 
that the vibronic motion affects current-voltage spectra. On the other
hand, when the molecule is coupled {\em strongly} to the electrodes, the
vibrational effects in resonant transport are almost negligible. This
phenomenon has been clearly justified in the reference~\cite{dom2}.
In the present work we have investigated all the essential features of
electron transport in the limit of strong molecule-to-electrode coupling,
and accordingly, we neglect the vibronic effect.

Throughout our work we have neglected the intra- and inter-site Coulomb 
interactions as well as the effect of the leads, which we wish to consider
in our future studies. Another important assumption is the zero 
temperature approximation. Although the electron-phonon interaction becomes 
significant at non-zero temperature, but in our case, we have already 
considered the dephasing effect phenomenologically at every molecular 
sites, which takes into account the effect of dephasing of electrons 
without any exchange of energy between the electron and the external 
leads. Accordingly, we compute the current-voltage relation at absolute 
zero temperature. At finite non-zero temperature, the transmission peaks 
will broaden due to thermal broadening effect, but the basic features 
will not change significantly as long as the thermal energy $(k_{B}T)$ 
is less than the average level spacing of the benzene molecule. 
Now-a-days various organic compounds and polymers
(for example, Tris(8-hydroxyquinolino) aluminum (Alq$_3$) and 
1,2,3,4,5-pentaphenylcyclopentadiene (PPCP)) are used as
electroluminescent devices. Knowledge of electronic structure and
current-voltage relation is important to elucidate the light emitting
mechanism~\cite{sug}. It may be interesting to see the effect of 
electron dephasing on this mechanism. Our presented results may
be useful in understanding the effect of phase breaking processes 
on the basic two-terminal molecular transport phenomena.

\end{document}